\documentclass[12pt]{article}
\usepackage[dvips]{graphicx}

\newcommand{\be}{\begin{eqnarray} }
\newcommand{\ee}{\end{eqnarray} }
\newcommand{\beq}{\begin{equation} }
\newcommand{\eeq}{\end{equation} }

\newcommand{\simleq}{\; \raisebox{-0.4ex}{\tiny$\stackrel
{{\textstyle<}}{\sim}$}\;} 

\newcommand{\palka}{{\Bigl |}}

\newcommand{\jpsi}{{J/\psi}}

\begin{document}
\begin{center}
\large\bf
On duality of Drell-Yan and $J/\psi$ production processes 
\rm
 \vskip 1cm
{A. Sissakian,  O. Shevchenko, O. Ivanov\\
{\it  Joint Institute for Nuclear Research, 141980 Dubna, Russia}
}
\end{center}
\begin{abstract}
It is studied the model on $\jpsi$ production allowing to extract parton distribution functions (PDFs)
from the combined analysis with both data on
Drell-Yan and $\jpsi$ production processes. 
It is shown that 
this, so attractive from theoretical point of view, model, can be safely used in the low energy region $E\simleq100GeV$.
The significance of gluon contributions to the $\jpsi$ cross-sections is investigated.
The obtained results in the high energy region occur to be rather surprising.
\end{abstract}
\begin{flushleft}
\end{flushleft}

Nowadays we see the growing interest \cite{anselmino,barone} to the close analogy (duality) 
between Drell-Yan (DY) $H_1H_2\to \gamma^*X\to l^+l^-X$ and $J/\psi$ $H_1H_2\to\jpsi X\to l^+l^-X$ production mechanisms 
(see textbook \cite{leader} for review).
It is assumed that a such analogy/duality occurs at relatively low energies, when the gluon-gluon
fusion ($gg$) mechanism of $J/\psi$ production is dominated by the quark-antiquark fusion ($\bar qq$).
Then, since $\jpsi$ is a vector particle like $\gamma$ and the helicity structure of $\bar qq(\jpsi)$
and $(\bar qq)\gamma^*$ couplings is the same, one can get the $\jpsi$ production cross-section 
from the DY process cross-section applying the simple replacement 
%\footnote{Notice that 
%in Ref. \cite{leader} the analogous replacement is proposed not only for $\jpsi$ 
%but for any vector meson production. 
%}
\be
\label{replacement}
{16 \pi^2 \alpha^2 e_q^2 \to (g_q^\jpsi)^2 \, (g_{\ell}^\jpsi)^2 },\quad{
\frac{1}{M^4} \to \frac{1}{(M^2 - M_{J/\psi}^2)^2 +  
M_{J/\psi}^2  \Gamma_{J/\psi}^2} \>, 
}
\ee
where $M^2\equiv Q^2$ is the squared mass of dilepton pair, $M^2_{\jpsi}\simeq 9.59\,GeV^2$ is the squared $\jpsi$ mass
and $\Gamma_\jpsi$ is the full $\jpsi$ width. It is believed that the model (\ref{replacement})
can be applied in both unpolarized \cite{barone} and polarized \cite{anselmino} cases. The later
is due to the identical helicity and vector structure of $\gamma^*$ and $\jpsi$ elementary channels
(all $\gamma^\mu$ couplings). In particular, the model given by Eq. (\ref{replacement}) 
was exploited in Ref. \cite{anselmino} in the double-polarized (both hadrons in initial state are
transversely polarized) case. 

%\vskip 1cm
It should be noticed here, that though the pure vector $\gamma_\mu$ coupling (Dirac coupling)
is very likely in the $\jpsi$ production, still it is a hypothesis. In principle, still the effects
due to $\sigma_{\mu\nu}$ coupling (Pauli coupling) are possible. Later, for the number
of processes with the different {\it unpolarized} colliding hadrons it will be shown that 
the model (\ref{replacement}), based on pure vector coupling, indeed works well at low energies.
It can be considered as a strong argument in favor of pure vector coupling mechanism.
At the same time, in {\it polarized} case we still have no necessary data to check
the pure vector coupling hypothesis. Thus, for the comprehensive
verification of the ``duality'' model (\ref{replacement}) it is of 
crucial importance to collect the data on $\jpsi$ production with the polarized 
colliding hadrons. This is a strong additional motivation to perform the such experiments.
At present the $\jpsi$ production processes with polarized proton/deutron target are planned
at J-PARC \cite{jparc} (proton beam) and COMPASS \cite{compass} (pion beam). The respective experiment
with the polarized antiproton beam is possible at GSI (this possibility is now under discussion
 \cite{pax}).

%\vskip 1cm

The advantage of model (\ref{replacement}) is that in the region of $u$-quark dominance
(large Bjorken $x$) all couplings exactly cancel out in the ratios of cross-sections (like asymmetries),
so that they become absolutely the same for the  DY and $\jpsi$ production processes
(see Eq. (23) in Ref. \cite{anselmino} and Eq. (10) in Ref. \cite{barone}).
Thus, in this kinematical region it does not matter for the $u$ quark PDFs 
extraction where do the dilepton pair production
events come from: from continuum or from $\jpsi$ production region. 
Certainly, the such possibility to use $\jpsi$ production for PDFs extraction
is very attractive  because the dilepton production rate in the $\jpsi$ production region
is two orders of magnitude higher than in the continuum region above the $\jpsi$
mass. In particular, the model (\ref{replacement}) give us the possibility
to extract transversity $h_{1u}$ and the first moment of
Boer-Mulders function $h_{1u}^{\perp(1)}$ from $\jpsi$ production region. Namely,
in the large $x$ region 
the equations (Eqs. (19), (20) in Ref. \cite{our1} and Eqs. (12), (13) in Ref. \cite{our2}) 
for unpolarized $H_1H_2\to \gamma^*X\to l^+l^-X$ 
\be
\hat k=8\frac{h_{1\bar u}^{\perp(1)}(x_A)|_A h_{1 u}^{\perp(1)}(x_B)|_B+(u\leftrightarrow \bar u)}
{f_{1\bar u}(x_A)|_Af_{1 u}(x_B)|_B+(u\leftrightarrow\bar u)}
\ee
and single-polarized $H_1H_2^\uparrow\to \gamma^*X\to l^+l^-X$
\be
\hat A=-\frac{1}{2}\frac{h_{1\bar u}^{\perp(1)}(x_A)|_A h_{1 u}(x_{B^\uparrow})|_{B^\uparrow}+(u\leftrightarrow \bar u)}
{f_{1\bar u}(x_A)|_Af_{1u}(x_{B^\uparrow})|_{B^\uparrow}+(u\leftrightarrow\bar u)}
\ee
DY processes 
are the same for the respective $\jpsi$ production processes 
$H_1H_2\to\jpsi X\to l^+l^-X$ and $H_1H_2^\uparrow\to\jpsi X\to l^+l^-X$, providing us an access to $h_{1u}$
and $h_{1u}^{\perp(1)}$ as a result of combined analysis of DY and $\jpsi$ data.

Certainly, the ``duality'' model (\ref{replacement}) is applicable only in the such kinematical
regions where among the elementary processes contributing to $\jpsi$ production
the quark-antiquark fusion process dominates over the gluon-gluon fusion.
Qualitatively, it is clear that the gluon contributions should be suppressed at low
energies: at fixed $Q^2$ value ($9.6\,GeV^2$ for $\jpsi$ production we deal with) $x$ increase when $s\equiv (p_{H_1}+p_{H_2})^2$
decreases, while the ratio $g(x)/q(x)$ sharply decreases when $x$ increases.
However, we certainly need the quantitative analysis to find the region where
we can safely neglect the gluons. The main goal of this paper is to perform the such quantitative
analysis. To this end we, besides of the model (\ref{replacement}), will consider
the most popular and well established model on $\jpsi$ production which includes all 
elementary processes (quark-antiquark as well as
gluon-gluon fusion) and believed to be valid at any energies of incident
hadrons -- see \cite{vogt} and references therein. Then we will compare the predictions of
both these models with the respective existing experimental data \cite{corden}-\cite{prl42}.

To cancel unknown constants, we will study not absolute cross-sections
but the ratios\footnote{Let us recall that as a final goal we are interested namely in
the ratios -- asymmetries, which can give us an access to the different PDFs.} 
of cross-sections corresponding to the variety of hadrons/nucleus in initial state.
Namely, we will consider the ratios of the angle and $x_F$ integrated   (the later in the forward hemisphere $x_F>0$)
cross-sections on $\jpsi$ production $\sigma_{pp}/\sigma_{\pi^{\pm}p}$, $\sigma_{pA}/\sigma_{\pi^{\pm}A}$ 
and $\sigma_{pp}/\sigma_{\bar pp}$,
where the symbol $A$ denotes the different target nucleus (these are $W$, $C$, $Ca$, $Cu$, $Pt$ here).
The point is that the hadrons $\pi^\pm$ and $\bar p$ (on the contrary to proton and nuclei) contain the antiquark in the valence state.
That is why it is very important to study the ratios like $\sigma_{pp}/\sigma_{\pi^{\pm}p}$ 
and $\sigma_{pp}/\sigma_{\bar pp}$
which should possess very specific behavior: they should sharply decrease when $s$ decrease (large $x$, valence quark/antiquark
dominance region) and they should increase up to unity when $s$ takes large values (small $x$, sea quark and gluon dominance region).
The most of data on $\jpsi$ production giving us an access to the such kind of ratios  are the data in the forward hemisphere
with the pion or antiproton beams colliding with the proton or nuclear targets \cite{corden}-\cite{prl42}.

The simple ``duality'' model (\ref{replacement}) applied to the ratio of cross-sections $[\sigma_{H_1H_2}/\sigma_{H'_1H_2}]_{x_F>0}$
gives
\be
\label{sigdy}
[\sigma_{H_1H_2}/\sigma_{H'_1H_2}]_{x_F>0}=\frac{\int_0^{1-m_{\jpsi}^2/s}dx_F[s(x_1+x_2)]^{-1} F_{q\bar q}^{H_1H_2}}{\int_0^{1-m_{\jpsi}^2/s}dx_F[s(x_1+x_2)]^{-1}F_{q\bar q}^{H'_1H_2}}, 
\ee
where the quark/antiquark flow $F_{q\bar q}^{H_1H_2}$ is given by 
\be
\label{flowdy}
F_{q\bar q}^{H_1H_2}=\sum_{q=u,d,s}[q^{H_1}(x_{1})\bar q^{H_2}(x_{2})+\bar q^{H_1}(x_{1})q^{H_{2}}(x_{2})],
\ee
and $x_{{1,2}}$ are expressed via $x_F=x_{1}-x_{2}$ as  $x_{{1,2}}=[\pm x_{F}+\sqrt{x_F^2+4m^2_{\jpsi}/s}]/2$.
Notice that obtaining Eqs. (\ref{sigdy}), (\ref{flowdy}) we put the ratios $g_d^\jpsi/g_u^\jpsi$ and $g_s^\jpsi/g_u^\jpsi$
equal to unity in accordance with the existing experimental data \cite{prl42Ca}. Indeed, the data \cite{prl42Ca} on $\jpsi$ production
with the absolutely symmetric (so that the cross section per nucleon reads 
$\sigma_{\pi^\pm C}=[\sigma_{\pi^\pm p}+\sigma_{\pi^\pm n}]/2$) carbon target  clearly indicates that
the ratio $\sigma_{\pi^+ C}/\sigma_{\pi^- C}$ is seem to be near unity in $\jpsi$ region, while it falls toward $1/4$ above the dilepton
pair mass -- see Fig. 2 in Ref. \cite{prl42Ca} and discussion around. 
This is a good argument that on the contrary to $q\bar q$ annihilation mechanism for DY process (where
$d$ quark is suppressed by the factor $e_d^2/e_u^2=1/4$), for $\jpsi$ production $u$ and $d$ quarks should enter the cross-sections
with the same charge factors: $g_d^\jpsi/g_u^\jpsi\simeq1$. We also use the analogous relation $g_s^\jpsi/g_u^\jpsi\simeq1$
keeping in mind that the squared sea strange quark contribution is not so significant.

The main difference of the ``gluon evaporation'' model from the model (\ref{replacement}) is that the former in 
addition to $q\bar q$ fusion contribution $F_{qq}^{H_1H_2}$ given by Eq. (\ref{flowdy}) contains also 
gluon-gluon fusion contribution $F_{gg}^{H_1H_2}=g^{H_1}(x_{1})g^{H_2}(x_{2})$.
Applied to the ratios $[\sigma_{H_1H_2}/\sigma_{H'_1H_2}]_{x_F>0}$, the ``gluon evaporation'' model produces: 
\be
\label{evap1}
 &  & \frac{\sigma_{H_1H_2}|_{x_F>0}}{\sigma_{H_1'H_2}|_{x_F>0}}
=\frac{(\sigma_{q\bar q}+\sigma_{gg})_{H_1H_2}|_{x_F>0}}{(\sigma_{q\bar q}+\sigma_{gg})_{H'_1H_2}|_{x_F>0}},\\
 &  & \label{evap2}
 \sigma^{H_1H_2}_{q\bar q (gg)}\palka_{x_F>0}=\int_{4m_c^2}^{4m_D^2}dQ^2\int_0^{1-\frac{Q^2}{s}} dx_F\sigma^{q\bar q\to c\bar c(gg\to c\bar c)}(Q^2)\frac{x_1x_2}{Q^2(x_1+x_2)}F_{q\bar q(gg)}^{H_1H_2},
\ee
%\be
%\frac{d^2\sigma/dx_F\palka_{(H_1H_2\to J/\psi\to l^+l^-)}}{d^2\sigma/dx_F\palka_{(A'B'\to J/\psi\to l^+l^-)}}
%=\frac{d^2(\sigma_{q\bar q}+\sigma_{gg})/dx_F\palka_{(H_1H_2\to J/\psi\to l^+l^-)}}{d^2(\sigma_{q\bar q}+\sigma_{gg})/dx_F\palka_{(A'B'\to J/\psi\to l^+l^-)}},\\
%d\sigma^{H_1H_2}_{q\bar q (gg)}/{dx_F}=\int_{4m_c^2}^{4m_d^2}dQ^2\sigma^{q\bar q\to c\bar c}(Q^2)\frac{x_Ax_B}{Q^2(x_A+x_B)}H_{q\bar q(gg)}^{H_1H_2},
%\ee
where $2m_c=3.0\,GeV$ and $2m_D=3.74\,GeV$  are respectively the $c\bar c$ and open charm thresholds, 
while the elementary cross-sections $\sigma^{q\bar q\to c\bar c}$, $\sigma^{gg\to c\bar c}$ are proportional to $\alpha_s(Q^2)$ and can be found,
for example, in Ref. \cite{vogt} (see Eqs. (3) and (4)). For the comparison purposes we will consider also 
the ``gluon evaporation'' model without gluon contribution $F_{gg}^{H_1H_2}$. It is obvious that it differs from the ``duality''
model only by the extra integration over $Q^2$ with the weight $\sigma^{q\bar q\to c\bar c}(Q^2)$.

Let us first consider the ratios $\sigma_{pp}/\sigma_{\pi^\pm p}$.
The results obtained within the ``duality'' and ``gluon evaporation'' models are presented in Fig.1 % \ref{fig1}
in comparison with experimental data. First of all, one can conclude  that 
in the low energy region, near the first experimental point $\sqrt{s}\simeq8.7\,GeV$,
the curves corresponding to ``duality'' model and to ``gluon evaporation'' model with and without gluons almost merge
and equally well describe the existing experimental data. This is not suprising and is in agreement
with the qualitative predictions:   as it was discussed above, the gluon contribution should be suppressed in the low energy region. 
%Now, after we performed the respective quantitative 
%analysis, o
At the same time, the results in the high energy region occur to be rather surprising: 
even at very high energies (150 GeV and 200 GeV)
the gluon contribution seems to be insignificant in the ratios  $\sigma_{pp}/\sigma_{\pi^\pm p}$
and the curves with and without gluon contributions $F_{gg}^{H_1H_2}$ equally well describe the existing data.
Thus, in the case of 150$\div$200 GeV pion beam and the
proton target we need to improve the quality of data on $\jpsi$ production
to distinguish between the models with and without gluon contribution. The such measurements can be performed, for example,
by COMPASS collaboration \cite{compass}, where the possibility to study Drell-Yan and $\jpsi$ 
processes with the pion beam is now under consideration.

The absolutely analogous picture, insignificance of the gluon contribution even at high energies,
occur also for the ratios $\sigma_{pA}/\sigma_{\pi^\pm A}$ with the different target nuclei -- see Fig.2. % \ref{fig2}. 
In Fig.2 % \ref{fig2} 
the data with the approximately equal ratio $Z/A\simeq 0.4$ are collected.
The curves corresponding to the model calculations are obtained with $Z/A=0.4$ and neglecting\footnote{Usually \cite{vogt} the nuclear effects
are accumulated in the multiplier $A^{\alpha}$. However, for $x_F$ integrated cross-sections
these factors differ a little from unity -- see Ref. \cite{vogt} and references therein.} the nuclear effects,
so that the cross-section per nucleon reads: 
$
\sigma_{hA}|_{h=\pi^\pm,p}=\frac{Z}{A}\sigma_{hp}+\left(1-\frac{Z}{A}\right)\sigma_{hn}.
$

Let us now consider the ratios $\sigma_{pp}/\sigma_{\bar pp}$ with the incident antiproton instead of pion. The results are presented
in Fig.3. % \ref{fig3}. 
While in the low energy region we again see the good agreement between the models with and without gluons
and the data (as it should be from the qualitative consideration), the situation in high energy region is absolutely different.
First, one can conclude, that the gluon contribution becomes very significant in this kinematical region.
%\vskip 1cm
Second, and rather surprising conclusion, is that the widely used ``gluon evaporation'' model works
rather bad in this case -- the respective curve lies well below the data (bold solid line in
Fig. 3). Notice that this result
is in a strong disagreement with the statement made in Ref. \cite{plb252}, 
where the same experimental points were used.
Indeed, calculations within the ``gluon evaporation'' model presented in Ref. \cite{plb252} (bold dashed line in Fig.3) 
are in a good
agreement with the data points which was claimed there as a strong argument in favor of that model 
(see Fig.6 in Ref. \cite{plb252} and discussion around).
The reason of this disagreement is in the gluon\footnote{It is clear seen from Fig. 3. Indeed,
the curves corresponding to the
``gluon evaporation'' model without gluons, for both  old and modern parametrizations (thin dashed and solid
lines in Fig. 3),
practically merge.} sector of the model,
because the main difference of the parametrization \cite{DO} used in Ref. \cite{plb252} 
and the modern parametrizations\footnote{We present here the result (bold solid line in Fig. 3) with 
the modern and widely used GRV98 \cite{grv98} parametrization.
However, our calculations with another modern parametrizations produce 
the same picture (the results
differ a little from the respective results with the GRV98 parametrization).} used in our calculations 
is just the gluon distribution values. 
Certainly, one should trust namely the calculation where the modern parametrizations are used,
because $G(x)$ is much better fixed there by a lot of new data appeared since
the paper \cite{DO} issue. The respective results occur to be in the strong disagreement
with the data and, thus, 
one can conclude that to pass this test the gluon sector in the ``gluon
evaporation'' model should be essentially modified. Besides, all existing
nowadays models on $\jpsi$ production should also pass this test on high energy
behavior, and this is a subject of our future investigation. 
%\vskip 1cm

In conclusion, we have tested the ``duality'' model as well as ``gluon evaporation'' model in the different energy ranges.
It is shown that the ``duality'' model (as well as ``gluon evaporation'' model) works well in the low energy sector, 
$s\simleq 100GeV^2$. In this region the curves with and without gluon contributions almost coincide and equally well
describe the data. Thus, we can safely apply in this kinematical region so attractive from the theoretical point of view
``duality'' model. This give us the unique possibility to use the $\jpsi$ production region for combined
analysis with the data on Drell-Yan processes to extract PDFs we are interested in. This is an excellent 
possibility to essentially increase the statistics since the $\jpsi$ production cross-sections are in about two orders
of magnitude higher than the respective DY cross-sections.

On the other hand, we got two surprises in the high energy region. First is for the pion-proton/nuclei collisions,
where the gluon contribution seems to be insignificant even at the energies of incident pion about 150$\div$200 GeV.
The second surprise occurs for the antiproton-proton collisions. Here, on the one hand,
the gluon contribution in the ratio $\sigma_{pp}/\sigma_{\bar pp}$  is very essential. On the other hand,
it is not properly described by the popular and widely used ``gluon evaporation'' model, when we use the
modern parametrizations on $G(x)$ instead of old (and rather incorrect) one. Thus, it seems that the ``gluon evaporation''
model should be properly modified.

Thus, the results of performed tests definitely show that both theoretical and experimental efforts 
are still necessary to answer the raised questions. We need the new information from both low and high energy regions.
In this respect it should be noticed that at present the new experiments with the proton-proton collisions
on Drell-Yan and $\jpsi$ processes are planned \cite{jparc} at J-PARC facility. The respective measurements with the  
low energy 
antiproton beam could be performed at GSI (see Ref. \cite{pax}). At the same time, the researches
on Drell-Yan and $\jpsi$ physics
at high energies (150$\div$200 GeV energy of incident pion) are now planned at COMPASS experiment \cite{compass}.

  The authors are grateful to M.~Anselmino, R.~Bertini,
 O.~Denisov, Y.~Goto, A.~Efremov, T.~Iwata, S.~Kazutaka, S.~Kumano, 
  A.~Maggiora, A.~Nagaytsev,  A.~Olshevsky, 
G.~Piragino, G.~Pontecorvo, S.~Sawada, I.~Savin, O.~Teryaev 
 for fruitful discussions. 
 The work  of O.S. and O.I. was supported by the Russian Foundation
 for Basic Research (project no. 05-02-17748).
O.S. thanks for hospitality RIKEN (Japan), and Yamagata University (Japan),
 where the part of this work was done.

\vfill\eject

\begin{figure}
\begin{tabular}{cc}
\includegraphics[height=6cm]{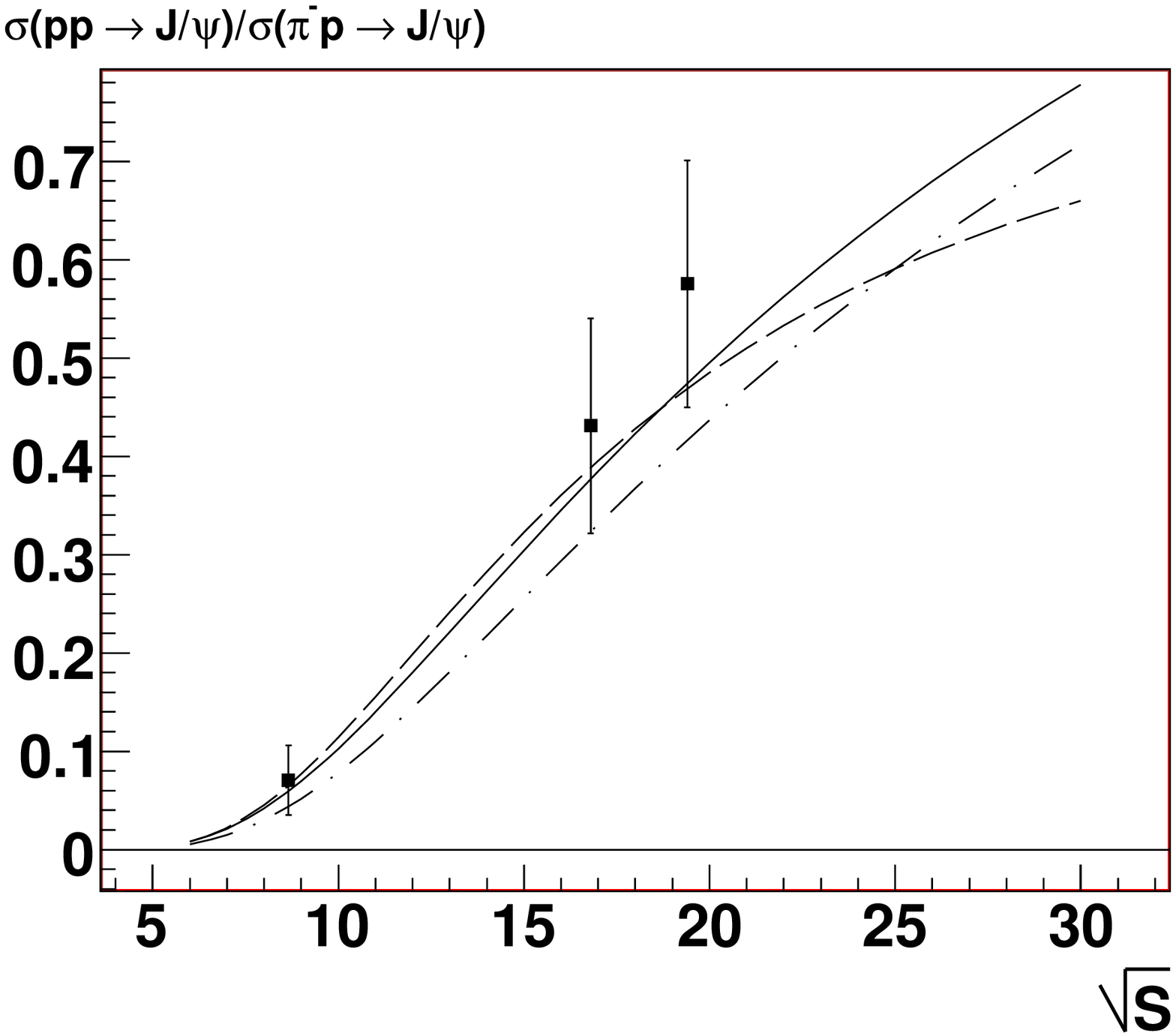} & 
\includegraphics[height=6cm]{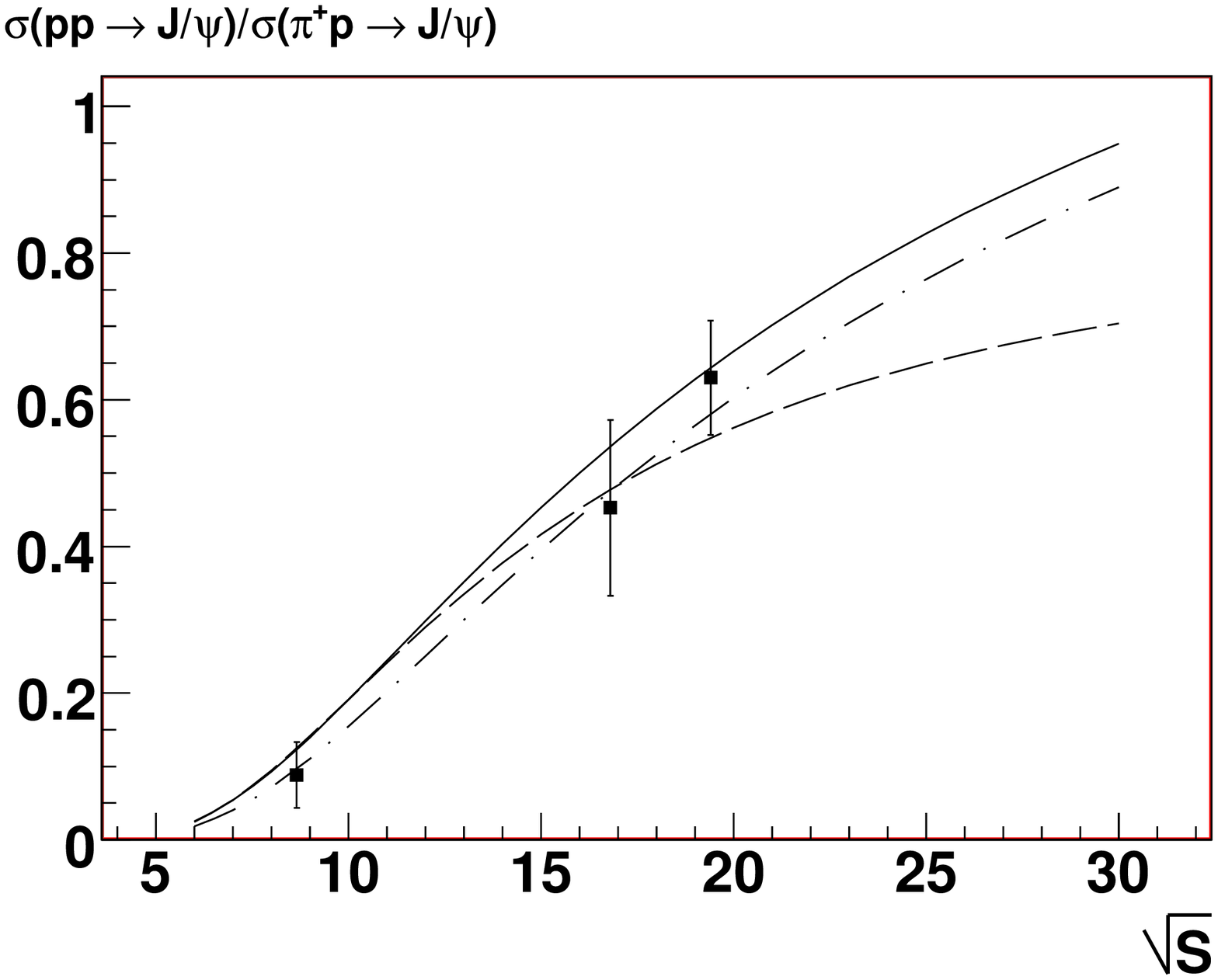}
\end{tabular}
\caption
{
Ratios of cross-sections $\sigma_{pp}/\sigma_{\pi^{\mp}p}$ on $J/\psi$ production 
calculated with two models in comparison with the experimental data.
Solid line corresponds to the ``duality'' model, Eqs. (\ref{sigdy}), (\ref{flowdy}).
Dashed line corresponds to the ``gluon evaporation'' model, Eqs. (\ref{evap1}), (\ref{evap2}).
Dot-dashed line corresponds to ``gluon evaporation'' model
without gluon contribution.
Here GRV94 \cite{grv94} parametrization for the proton PDFs
and GRV \cite{grvp} parametrization for the pion PDFs are used.
The experimental data (points with error bars) are taken from \cite{vogt} (Tables 2 and 3).
}
\label{fig1}
\end{figure}

\begin{figure}[b!]
\begin{tabular}{cc}
\includegraphics[height=6cm]{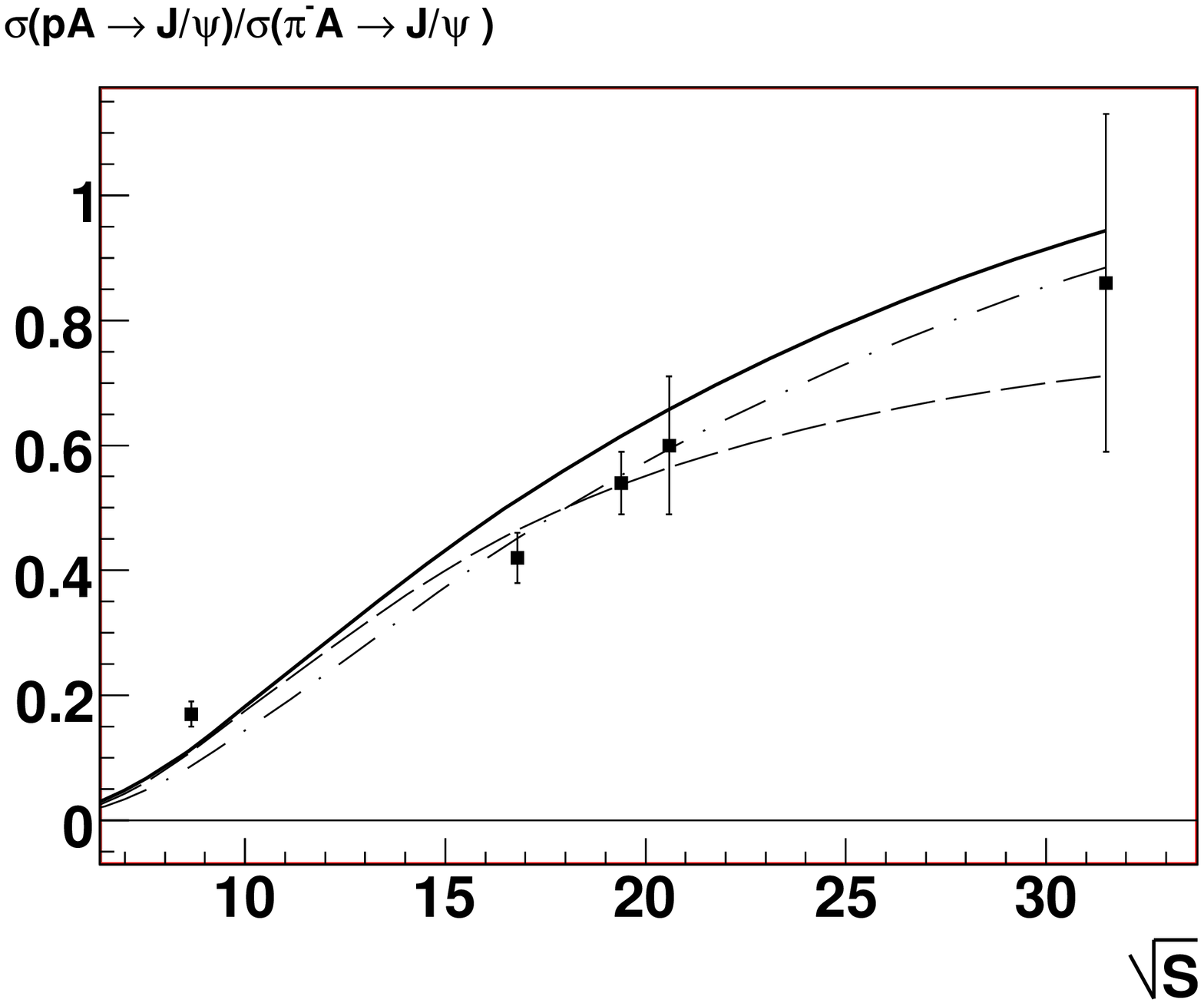} 
& 
\includegraphics[height=6cm]{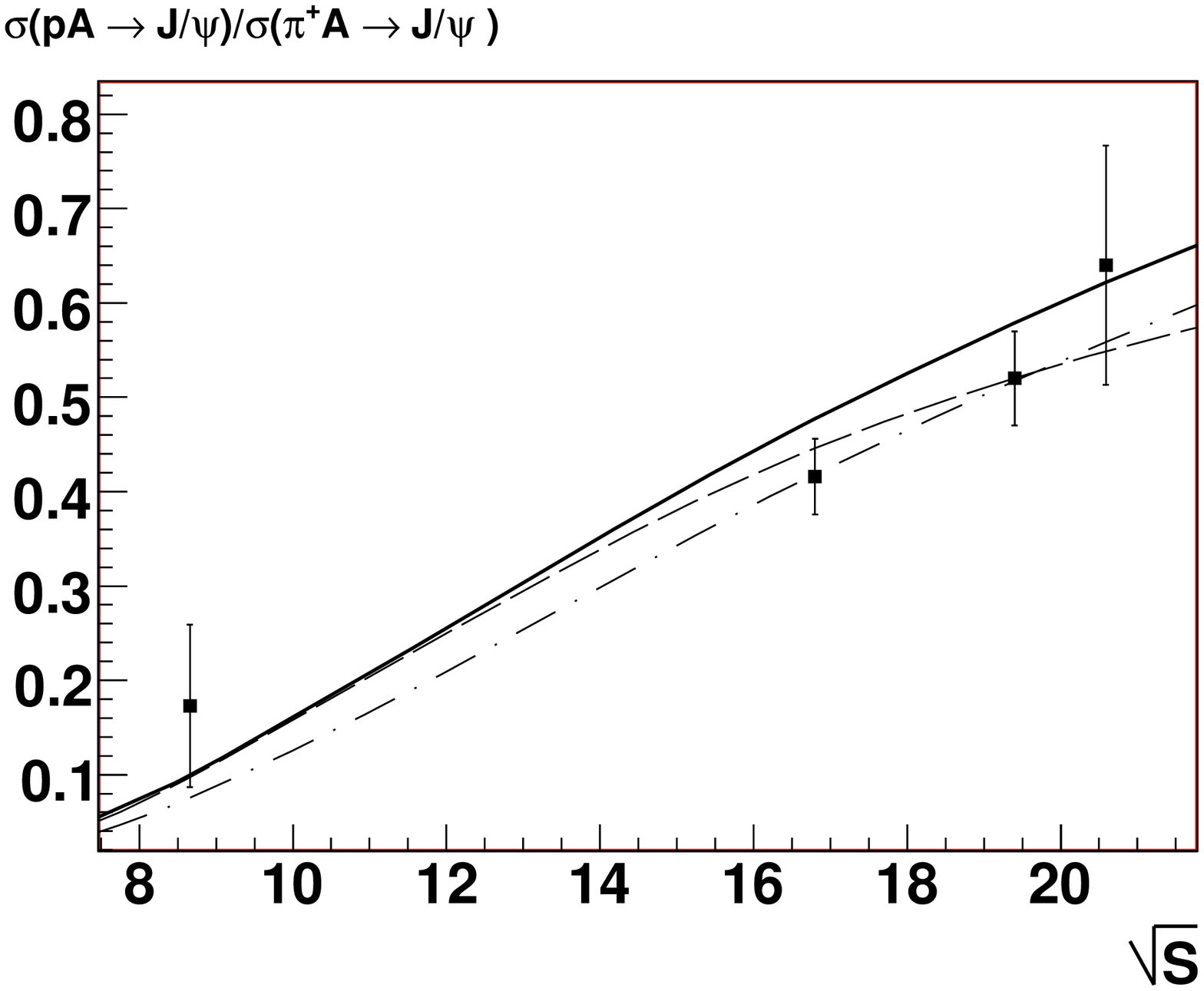}
\end{tabular}
\caption{
Ratios of cross-sections $\sigma_{pA}/\sigma_{\pi^\mp A}$ on $\jpsi$ production for the 
nuclear targets ($Z/A\simeq0.4$) calculated with two models
in comparison with the experimental data.
Solid line corresponds to the ``duality'' model, Eqs. (\ref{sigdy}), (\ref{flowdy}).
Dashed line corresponds to the ``gluon evaporation model'', Eqs. (\ref{evap1}), (\ref{evap2}).
Dot-dashed line corresponds to the ``gluon evaporation'' model
without gluon contributions.
Here GRV94 \cite{grv94} parametrization for proton PDFs
and GRV \cite{grvp} parametrization for pion PDFs are used.
The points with error bars correspond to experimental data.
First point: W, Z/A=0.40 \cite{cordenW}; second and third points: Pt, Z/A=0.40 \cite{badier};
fourth point: C, Z/A=0.5 \cite{prl42};
fifth point: Be, Z/A=0.44 \cite{abramov}.
}
\label{fig2}
\end{figure}

\begin{figure}
\begin{tabular}{cc}
\includegraphics[height=9cm]{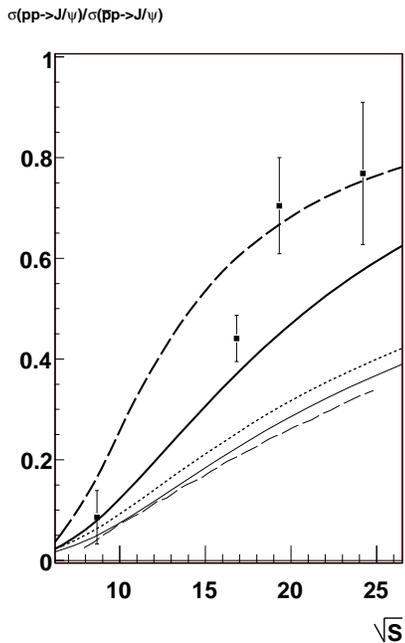} 
\end{tabular}
\caption{
Ratios of cross-sections $\sigma_{pp}/\sigma_{\bar pp}$ on $\jpsi$
production for the proton target calculated with two models in comparison with
experimental data. The experimental data are taken from Ref. \cite{plb252},
where the respective data of NA3, WA39 and UA6 collaboations were collected
(see Fig.6 in Ref. \cite{plb252}).  The bold solid and dashed lines correspond to
calculation with the ``gluon evaporation'' model (with
GRV98 \cite{grv98} and  Duke-Owens \cite{DO} parametrizations, respectively).
The thin solid and dashed lines correspond to calculations with the ``gluon
evaporation'' model without gluon contrubutions  (with
GRV98 \cite{grv98} and  Duke-Owens \cite{DO} parametrizations, respectively). The dotted line correspond to
calculation with the ``duality'' model and GRV98 \cite{grv98} parametrization.
}
\label{fig3}
\end{figure}

\end{document}